\documentclass[iop, apj, tighten]{emulateapj}
\usepackage{amsmath,amsthm,amssymb,bm}

\shorttitle{Test-particle acceleration in 3D-RMHD}
\shortauthors{Dalena et al.}

\begin{document}

\title{Test-particle acceleration in a hierarchical three-dimensional turbulence model}

\author{S. Dalena$^{1,2,\dagger}$, A.~F. Rappazzo$^1$, P. Dmitruk$^3$, A. Greco$^2$, and W.~H. Matthaeus$^1$}
\affil{$^1$Bartol Research Institute, Department of Physics and Astronomy, 
University of Delaware, Delaware 19716, USA\\
$^2$Dipartimento di Fisica, Universit\`a della Calabria, I-87036 Cosenza, Italy\\
$^3$Departamento de Fisica, Facultad de Ciencias Exactas y Naturales, Universidad de Buenos Aires, Ciudad Universitaria, 1428 Buenos Aires, Argentina}
\email{$^{\dagger}$dalena@bartol.udel.edu, serena.dalena@fis.unical.it}

\begin{abstract}
The acceleration of charged particles is relevant to the solar corona over a broad range of scales and energies.  
High-energy particles are usually detected in concomitance with large energy release events like solar eruptions and flares, 
nevertheless acceleration can occur at smaller scales, characterized by dynamical activity near current sheets.  
To gain insight into the complex scenario of coronal charged particle acceleration, 
we investigate the properties of acceleration with a test-particle approach using 
three-dimensional magnetohydrodynamic (MHD) models.
These are obtained from direct solutions of the reduced MHD equations, well suited for a plasma embedded in a strong 
axial magnetic field, relevant to the inner heliosphere. 
A multi-box, multi-scale technique is used to solve the equations of motion for protons. 
This method allows us to resolve an extended range of scales present in the system,
namely from the ion inertial scale of the order of a meter up to macroscopic scales
of the order of $10$\,km ($1/100$th of the outer scale of the system). 
This new technique is useful to identify the mechanisms that, acting at different scales, 
are responsible for acceleration to high energies of a small fraction of the particles in the coronal plasma.
We report results that describe acceleration at different stages over a broad range of time, length and energy scales.
\end{abstract}


\section{Introduction}

Understanding the mechanisms that accelerate charged particles
in dynamical plasmas is an intense subject of research, with
applications to astrophysical, space, and laboratory environments \citep{cvb12}.
Charged particles are energized in solar flares, planetary
magnetospheres, at interplanetary shocks, and in the interstellar medium.
In the solar corona electrons can be accelerated to tens of MeV, 
while ions gain energies up to several GeV, 
and high-energy particles may carry up to $30 \%$ of the
energy budget of a flare \cite[for a review see][]{asc02, km07}.
The fast release of energy in solar flares is commonly attributed to magnetic
reconnection events. 
In such complex natural systems, several processes may contribute to 
the observed particle energization, and these may be distinguished in part 
by the locations and scales at which they occur.
In this way the resulting acceleration might be viewed as a \textit{multi-stage process}
\cite[see, e.g.,][]{jag86, gor05, bro09}.

In particular the role of scattering from magnetic irregularities
is well known to be an important factor \citep{Jokipii66}:
particles can be accelerated by the large scale {\bf electric} fields 
\citep{Speiser65, Litvinenko96}, as well as by the 
smaller scales of reconnecting fields \citep{DrakeEA05} and/or 
discontinuities, including acceleration in shocks 
\citep{Giacalone&Jokipii96, Giacalone05}.
Furthermore they are also accelerated by resonances with structures at
various scales, and acceleration mechanisms may be strongly influenced
in various contexts by changes of magnetic connectivity \citep[e.g.,][]{rmr12}.
Thus essentially all length scales may be important in the investigation of 
the energization process, giving rise to a rather complex behavior in the case of
turbulent fields, characterized by the presence of energy at all scales from 
large to small and of current and vorticity coherent structures, that
can cause resonant or nonresonant interactions \citep{MatthaeusEA84, KobakEA00}.
Indeed, although the smooth or average electromagnetic fields are associated with 
a variety of possible particle motions \citep{Speiser65, Sonnerup71, Cowley78},
magnetic fluctuations can strongly influence both the local electric field and particle motion.
For example strong resonant pitch angle scattering, as
well as the formation of transient trapping centers, such as secondary islands, are 
directly associated with the presence of electromagnetic fluctuations \citep{MatthaeusLamkin86, RuffoloEA03, DrakeEA06}.

Such effects are expected to be widespread since turbulent magnetic fluctuations are observed in space 
plasmas in practically  all environments. Fully developed turbulence involves itself a hierarchical process 
in which many scales of motion are engaged in the energy transfer.
As particles become energized, their interactions with the turbulence sample a correspondingly wide range of scales.
From an observational point of view, different spectra and intensity variations 
have been detected from sources of nonthermal radiation in many flares \cite[see, e.g.,][]{tak95, ems03, fle07}, 
which is also hard to explain invoking only a single acceleration mechanism. 

Because of the complexity of particle motion, a useful approach has been to
resort to numerical experiments based on test particle simulations, in which no interaction among 
the particles and no back reaction to the imposed electromagnetic fields is considered
\citep{LitvinenkoEA93, TurkmaniEA05, DallaEA05, OnofriEA06}.
However, processes that span a wide range of scales pose profound challenges because 
the different scales require different time and spatial resolutions, 
and qualitatively different phenomenologies may apply.
For instance, in the description of a coronal loop the typical scale of the system 
$L$ measures thousands of kilometers.
In contrast the ion inertial scale $d_i$, the typical scale at which dissipative effects become important \citep{SmithEA01}, 
is of the order of a meter and is comparable with the typical gyroradius, 
$\rho = v_\textrm{th}/\Omega$, of a particle moving 
at the thermal speed $v_\textrm{th}$ with a gyrofrequency $\Omega$.

Such conditions, spanning a wide range of length scales, are typical in space plasmas.
For example, in the solar wind it can be estimated that $L/d_i \sim 10^5$.
In a coronal hole (at about one solar radius) and in magnetically confined
regions of the solar corona $L/d_i \sim 10^6$.
Attempting to resolve this full range of scales requires an enormous cost in
terms of computing power at present standards.
In galactic astrophysics even larger $L/d_i$ may be relevant. 
To attack such problems, approximate models, such as the guiding center and gyrokinetics, 
have been employed to describe particle motion in a strong magnetic 
field \cite[see, e.g.,][]{lee83, bv01, giu05, gor05, LeheEA09}.
These models are derived using magnetic moment $\mu$ conservation, 
so their validity is limited to specific circumstances.
In fact the equation of motion of the particle guiding center can be solved in the case of small 
gyroradius and large gyrofrequency, $\rho \ll l$ and $\Omega \gg T^{-1}$, where $l$ and $T$ 
should be interpreted as the typical length and time scales of the {\it local} electromagnetic field.
As a consequence, even if particles are strongly magnetized globally, 
non-adiabatic behavior may develop at local scales \citep{bro09, dal12a},
in particular in the presence of sharp field gradients, near reconnection sites 
or in  presence of well developed magnetohydrodynamic turbulence. 
In these situations the presence of resonant effects and/or coherent structures could violate 
magnetic moment $\mu$ conservation. 
\begin{figure}
\begin{center}
\includegraphics[width=8.5cm]{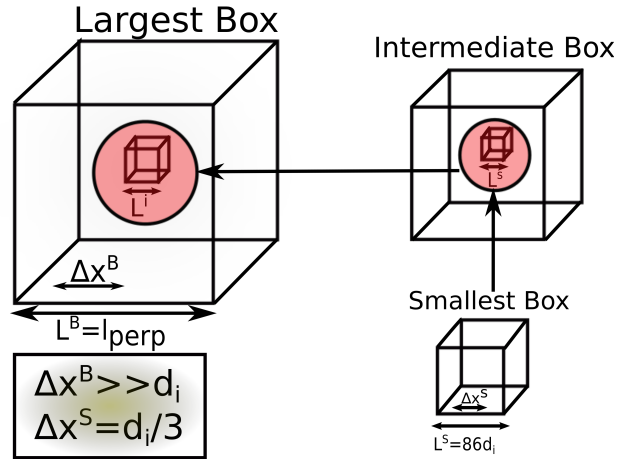}
\end{center}
\caption{Schematic representation of the hierarchical computational 
boxes used in the multi-scale model.
The boxes are progressively larger to accommodate higher energy particles
(with correspondingly larger giroradii). 
Physical fields in a larger box are obtained rescaling those from the smaller box.
Indicating, e.g.,  with $L$ and $\Delta x$ respectively the linear length
and grid resolution along the $x$-direction of the smaller box, the fields in the
larger box of linear length $L'$ and grid size $\Delta x' = \Delta x\, L'/L$
are built rescaling the fields from the smaller box by a factor $L'/L$ in each
linear direction and remapping them to the bigger box.}

\label{fig:box}
\end{figure}

To represent turbulent fields at all relevant scales, one approach is to use gridless methods
with prescribed analytical fields \cite[see, e.g.,][]{Giacalone&Jokipii96, qin02, byk08, fras08, fras12}.
In principle, this method can represent a wide range of scales by correspondingly increasing the number of modes
retained in the representation. However, the direct gridless implementation becomes extremely
onerous in that case. These models also most often employ random phases, which prevent
representation of coherent small scale structures, such as current sheets and vorticity
concentrations, that can play a key role in particle acceleration.

Therefore we have developed a new multi-box, multi-scale test-particle technique
to resolve the wide range of time and length scales usually involved in the description 
of space and astrophysical plasmas. The main idea is to study particles dynamics at different steps during their evolution,
moving from several simulation boxes of different dimensions (for a schematic picture see Figure~\ref{fig:box}). 
The nonrelativistic equations of motion of a particle moving in an electromagnetic 
field are solved numerically, keeping approximately constant the resolution of the particle 
{\bf gyroradius} $\rho$ going from a smaller to a larger box. 
This has been accomplished choosing for every box a grid scale $\Delta x$ that satisfies the condition $\Delta x < \rho$. 
Different spatial and temporal resolutions in different regions allow us to treat with the same
accuracy all the scales involved in the system.
In this paper we present this new numerical technique applied to the study of particle acceleration in a coronal loop. 
We report results that describe acceleration in direct simulations 
of reduced magnetohydrodynamic (RMHD) turbulence at 
different stages, over a broad range of time, length and energy scales.

The paper is organized as follows.
In Section~\ref{sec:2} we describe the equations and properties of the MHD fields and of the particles.
In Section~\ref{sec:3} the multi-box technique is introduced.
In Section~\ref{sec:4} results are given for energization 
of protons obeying the nonrelativistic equations of motion.
Section~\ref{sec:5} contains our conclusions.

\section{Methods} \label{sec:2}

The procedure consists of a reduced Magnetohydrodynamics (RMHD) simulation, 
appropriate for the low frequency dynamics of a plasma in a strong externally supported 
magnetic field ${\bf B}_0 = B_0{\bf e}_z$, followed by integration of test particle orbits.
The RMHD equations are solved in a Cartesian box with aspect ratio $l_\parallel/l_\perp = 10$ 
(ratio of characteristic lengths parallel and perpendicular to ${\bf B}_0$),
and a numerical grid of $256 \times 256 \times 120$ points. 
For more details on the numerical methods and code, see \cite{rved07, rved08}.

Particle trajectories are integrated in a single snapshot of the RMHD field, taken at $t=t_{nl}=10t_A$,
where $t_{nl}$ is the nonlinear time and $t_A$ is the Alfv\'en crossing time. 
The nonlinear stage is characterized by the presence of current
sheets elongated along the axial direction and represents a statistical steady state
in which energies fluctuate around a mean value, and on average total dissipation and Poynting flux are in balance.
The quasi-static approach in which magnetic and electric fields do not change with time is justified by assuming 
that the particle acceleration time is shorter than the characteristic time 
of MHD system evolution.

Another possible way to attack the multi-scale problem is to study the acceleration of 
particles moving in an analytic electromagnetic field \citep{TuEA93, GrayEA96, BieberEA96},
given as a Fourier series. In the limit of an infinite number of wave modes, 
the turbulence can approach a pure symmetry, and may be for example
isotropic and spatially homogeneous \citep{Batchelor60}. Methods that
sum a sparse set of plane waves with randomly distributed propagation direction, 
and with random polarizations and phases, may be referred to as gridless methods
\citep{Giacalone&Jokipii99}. These have been employed in studies of the transport 
of cosmic rays performing numerical simulations of test-particles interacting 
with a time stationary turbulent magnetic field. These methods typically use a
smaller number of Fourier modes due to the requirement of direct summation of
the series at many points. These methods also lack coherent structures, such as current sheets.

While it has its own limitations, the multi-box technique we introduce here 
does allow for fields that have coherent structures and, given the large number
of modes it permits, it allows for development of resonances and related diffusion processes 
that require a high density of modes. Furthermore, by rescaling the fluctuations
across multiple boxes, the multi-box technique permits the study of particle
acceleration including these important effects: coherent structures and resonances across
all relevant scales during the multi-stage acceleration process.

\subsection{Reduced MHD simulation}

A coronal loop is a closed magnetic structure threaded by a strong axial field, 
with the footpoints rooted in the photosphere.
This makes it a strongly anisotropic system, as measured by the relative 
magnitude of the Alfv\'en velocity $v_{_{\!A}} = 2000$~km\,s$^{-1}$
(associated to the strong axial field $B_0$)
compared to the typical photospheric velocity $v_\textrm{ph}\sim1$~km\,s$^{-1}$.
These relatively slow motions shuffle the footpoints of the axially elongated 
magnetic field lines giving rise to a coronal magnetic field $b$ small compared
to the guide field $\delta b/B_0 \ll 1$, and the plasma dynamics are then
expected to develop in the reduced MHD regime
\citep{str76} (for a more detailed analysis see \cite{evpp96, dg97}).
The loop dynamics may be studied in a simplified geometry, neglecting any 
curvature effect, as a ``straightened out'' Cartesian box, 
with an orthogonal square cross section of size $L = l_\perp$, and an axial
length $l_z$ embedded in an axial homogeneous uniform magnetic field 
${\bf B}_0 = B_0{\bf e}_z$.

The dynamics are integrated with the (nondimensional) equations of reduced magnetohydrodynamics (RMHD) \citep{kp74,str76,mon82},
well suited for a plasma embedded in a strong axial magnetic field.
In dimensionless form they are given by:
{\setlength\arraycolsep{-10pt}
\begin{eqnarray}
&& \partial_t \mathbf{u}  + 
 \mathbf{u} \cdot \nabla  \mathbf{u} = 
- \nabla P + \mathbf{b} \cdot \nabla  \mathbf{b}
+ c_{_{\!A}} \partial_z \mathbf{b}
+ \frac{ 1 }{Re} \nabla^{2} \mathbf{u}, 
\label{eq:eq1} \\[.1em]
&&\partial_t \mathbf{b}  + 
\mathbf{u} \cdot \nabla \mathbf{b} = 
\mathbf{b} \cdot \nabla \mathbf{u} 
+ c_{_{\!A}} \partial_z \mathbf{u}
+ \frac{ 1 }{Re_m} \nabla^{2} \mathbf{b}, \\[1em]
&&
\nabla \cdot \mathbf{u} =  \nabla \cdot \mathbf{b} = 0.
\label{eq:eq2}
\end{eqnarray}
}%
Here, gradient and Laplacian operators have only transverse  (x--y) 
components as do velocity and magnetic field vectors 
($u_z = b_z = 0$), while $P$ is the total (plasma plus magnetic) pressure.
Incompressibility in RMHD equations follows from the large value of the axial magnetic field 
\citep{str76} and they remain valid also for low plasma $\beta$ systems \citep{zm92, bns98}
such as the corona, where the plasma $\beta \sim 0.01$ in closed regions.
The plasma is then assumed to have a uniform constant  density $\rho_0$.

To render the equations nondimensional, we have first expressed the
magnetic field as an Alfv\'en velocity
$[ b \rightarrow b/\sqrt{4\pi \rho_0} ]$, 
and then all velocities have been normalized to the velocity 
$u^{\ast} = 1\,$km\,s$^{-1}$, the order of magnitude of photospheric convective motions.
Lengths and times are expressed in units of the perpendicular length of 
the computational box $\ell^\ast = \ell_\perp$ and its related crossing time
$t^\ast = \ell^\ast / u^\ast$.	
As a result, the linear terms $\propto \partial_z$ are multiplied by the dimensionless 
Alfv\'en velocity $c_A = v_A / u^\ast$, 
where $v_A = B_0 / \sqrt{4\pi \rho_0}$ is the Alfv\'en velocity associated with 
the axial magnetic field.

We adopt a standard simplified diffusion model, in which both the magnetic resistivity 
$\eta_m$ and dynamic viscosity $\nu$ are constant and uniform. 
The kinetic and magnetic Reynolds numbers are then given by
\begin{equation}
Re = \frac{\rho_0\, \ell^\ast\, u^\ast}{\nu}, \qquad
Re_m = \frac{4 \pi\, \ell^\ast\, u^\ast}{\eta_m c^2},
\end{equation}
where $c$ is the speed of light, and numerically they are given the same value 
$Re = Re_m =400$.

We solve numerically equations (\ref{eq:eq1})-(\ref{eq:eq2}) written in terms 
of the potentials  in Fourier space, i.e., we
advance the Fourier components in the $x$- and $y$-directions of the 
magnetic field and velocity scalar potentials \citep[for a more detailed description of the numerical
code and methods, see][]{rved07, rved08}. 
Along the $z$-direction, no Fourier transform is 
performed so that we can impose nonperiodic boundary conditions, and a central second-order 
finite-difference scheme is used. In the $x$-$y$ plane, a Fourier pseudospectral method 
is implemented. Time is discretized with a third-order Runge-Kutta method. We use a 
computational box with an aspect ratio of 10, which spans $0 \le x, y \le 1$,
$0 \le z \le 10$, and implement a numerical grid 
with $256 \times 256 \times 120$ points.


As boundary conditions at the bottom and top plates ($z=0$ and $z=10$,
the \emph{photospheric-mimicking} surfaces), 
we impose two independent velocity patterns, 
intended to mimic photospheric motions, 
made up of large spatial scale projected convection cell flow patterns 
constant in time \citep[e.g., see Figures~1 and 2 in][]{rved08}. 
These are essentially large-scale disordered vortices
that stir the footpoints of the \emph{line-tied} axially elongated 
magnetic field lines.
Similarly to standard forced MHD turbulence simulations this
boundary forcing injects energy into the system giving rise 
to an MHD turbulence cascade dominated by magnetic energy
\citep{evpp96, dg97, dgm03, rved08}. In physical space
current sheets form on very fast ideal timescales \citep{rp13},
and are aligned to the strong axial magnetic field
\citep[e.g., see Figure 18 in][]{rved08}.

Given its quasi-2D properties (perpendicular magnetic and velocity
fluctuations, current sheets elongated in the axial direction,
and the presence of a strong guide field), 
RMHD turbulence is an adequate model in which to study
particles acceleration, serving as a first simplified model with
3D characteristics, in between the more simple
and more extensively studied 2D case and the more
complex and challenging fully 3D setup.
Furthermore RMHD represents a good model for the low
corona, largely characterized by the presence of a strong guide 
field, in both open and closed regions.

\subsection{Particle integration}
The behavior of a test particle in electromagnetic fields obtained from RMHD 
is described by its time dependent position ${\bf r}(t)$ and three-dimensional 
velocity ${\bf v}(t)$, advanced according to the non-dimensional equations:
\begin{eqnarray} \label{eq:ODE}
\frac{d{\bf r}}{dt}            &     =     & {\bf v}, \label{eq:ODE1} \\
\frac{d{\bf v}}{dt}           &     =     & \beta({\bf E} + {\bf v} \times {\bf B}). \label{eq:ODE2}
\end{eqnarray}
The non-dimensional electric field $\mathbf{E}$ is obtained through Ohm's law
(normalized with the field $E_0 = v_A B_0/c$),
from the RMHD simulation magnetic field ${\bf B}$, plasma velocity {\bf u}, 
and current density ${\bf j}$:
\begin{equation}
\label{eq:Efield}
{\bf E} = -\frac{1}{c} {\bf u} \times {\bf B} + \eta{\bf j}.
\end{equation}
The nondimensional resistivity $\eta = 1/Re_m$ is equal to $2.5\times 10^{-3}$.
To render the equations non-dimensional, we use as characteristic quantities 
the turbulence correlation length $\lambda_c= l_\perp/4 = 1000$~km, 
the Alfv\'en velocity $v_{_{\!A}} = 200$~km\, s$^{-1}$, and the Alfv\'en crossing time $t_{_{\!A}} = \lambda_c/v_{_{\!A}}$
\citep{dal12b}.

The dimensionless parameter $\beta=\Omega t_{_{\!A}}$ in Equation~(\ref{eq:ODE2}) couples particle and 
field spatial and temporal scales and provides a particularly useful means to relate numerical 
experiments to space and astrophysical plasmas. Indeed it can be rewritten as \citep{amb88, dal12a}:
\begin{equation} \label{eq:beta}
\beta = \frac{q}{m} \frac{\sqrt{4\pi\rho_0}}{c} \lambda_c = \frac{\lambda_c}{d_{i}},
\end{equation}
where $d_i = c/ \omega_i =(c^2 m_i / 4 \pi n_i q_i^2)^{1/2}$ is the ion inertial scale. Thus the $\beta$
parameter measures the range of scales involved in a particular system.
In general, in a turbulent collisionless plasma, the bandwidth of the inertial range
fluctuations may extend from large fluctuations near the correlation scale $\lambda_c$ to
small fluctuations near the ion inertial scale $d_i$. In this case one expects that $\beta \gg 1$.
For the typical values in our simulations $d_i \sim 2$~m, and $\beta = \lambda_c/d_i \simeq 5 \times10^5$.

Considering that the average value of the coronal temperature  $\sim 10^6\,$K
corresponds to velocity of $\sim 200$ km\, s$^{-1}$ for protons,
the non-relativistic approximations used in Equations~(\ref{eq:ODE1})-(\ref{eq:ODE2})
is adequate to investigate the acceleration processes up to values of 
$\gamma = 1/\sqrt{1-(v/c)^2} \sim 1.1$, corresponding to energies $\sim 0.1\,$GeV
for protons. The simulations presented here consider durations that allow
protons to gain at most energies of the order of $0.1\,$GeV.
In order to attain longer integration times the relativistic analog of 
Equations~(\ref{eq:ODE1})-(\ref{eq:ODE2}) should be used,
and this is left to future investigations.

\section{Multi-box technique}\label{sec:3}

The appearance of the large number $\beta$ in Equation~(\ref{eq:ODE2}) makes it apparent 
why it is challenging to integrate test particle evolution with parameters appropriate to space and astrophysical plasmas. 
The Equations~(\ref{eq:ODE1}) and (\ref{eq:ODE2}) are very stiff (involving greatly differing time scales), 
while, as some particles may accelerate until gyroradii become comparable to $L$, there
is a huge range of spatial scales involved. 
Therefore it is very difficult to generate numerically relevant magnetic fields that 
have structures across the required range of scales.  

Moreover, when dealing with numerical simulation of test-particles in electromagnetic field, 
one of the most important requirement is the resolution of the particle gyroradius $\rho$ 
during the integration. 
Indeed $\rho$ determines the minimum scale of a magnetic field structure capable of affecting 
the dynamic of the particle.
If the value of $\rho$ is smaller than the grid scale value $\Delta x$, 
the test-particle will not be able to sample the different structures of the field and it will 
move as if it were under the action of a constant field. 
The parameter $\beta$ is related to the value of the particle gyroradius by 
the parameter $\epsilon = \rho/\lambda_c = v/\Omega = (v/v_{_{\!A}})/\beta$ \citep{MaceEA00},
sometimes called the dimensionless particle rigidity.

The typical velocity of a particle traveling at the thermal speed in the solar corona is 
usually a fraction of the characteristic Alfv\'en speed $v_{_{\!A}}$, 
thus $\rho \le \lambda_c/\beta$. 
The wider is the range of length scales involved, the smaller the 
gyroradius will be in comparison with the largest scale present in the system.
For instance, with the choice of our parameters, $\epsilon \sim 9\times10^{-7}$. 
On the other hand the grid resolution scale $\Delta x = L/N$ of the simulation box is determined 
by the number of grid points $N$ and the box length $L$. 
The resolution of the particle gyroradius in this case would require a number of grid 
points of the order of $10^7$ in the $x$, $y$, and $z$
directions respectively, not achievable at the present standard.

Another subtle point in the description of test-particle motion is the potential for resonances
that can contribute significantly to acceleration.
If particles are injected with a gyroradius very much smaller than the outer scale of the system $\lambda_c$,
it becomes possible that the simulation will lack resonances. 
The dimensionless particle rigidity $\epsilon$ can be related to the bend-over wavenumber of the turbulent energy spectrum, 
$k_\textrm{bo} = 1/\lambda_c$, and the minimum resonant wavenumber, 
$k^{r}_\textrm{min} = 1/\rho$,  as $\epsilon = k_\textrm{bo}/k^{r}_\textrm{min}$.
When $\rho \gg \lambda_c$ particles experience all possible $k$-modes in few gyroperiods,
resonating with the energy  containing scale ($k^{r}_\textrm{min} \ll k_\textrm{bo}$). 
For lower energies the test particles resonate in the inertial range ($k^{r}_\textrm{min} \sim k_\textrm{bo}$). 
However, when $\rho \ll \lambda_c$, particles will require higher and higher wavenumbers to resonate with, 
but it is possible that the numerical representation may not have sufficient power in the high wavenumber 
magnetic energy spectrum to produce these resonant effects in the numerical experiment \cite[see, e.g., Figure 6 of][]{dal12a}.

To deal with the numerical issues discussed above, we have developed a multi-box, 
multi-scale test-particle approach to integrate numerically  the nonrelativistic equations of motion
of a particle in electromagnetic fields. 
Particle dynamics is studied at different stages during the energization process,
moving between simulation boxes of different dimensions and spatial resolutions,
each satisfying the condition $\Delta x < \rho$.
With this technique, we have been able to investigate particle motion and acceleration
at different length scales, retaining simultaneously at each step the resolution of the particle gyroradius. 
The method allows us to treat with the same accuracy lengths of the order of few meters 
(comparable with the ion inertial scale in the system) as well as lengths up to several 
thousand kilometers (the size of a coronal loop).
This new technique is useful to identify the mechanisms that, acting on different scales, 
are responsible for acceleration to high energies of a small fraction of the 
particles in the coronal plasma.

Numerical simulations in the different boxes are performed using the same 
realization of the turbulent magnetic, velocity
and electric fields obtained from the direct solution of RMHD equations.
However, moving from the smallest to the biggest box, it is important to establish 
a physically reasonable scaling of these fields to avoid anomalous behavior.
With this intention, we assume that the magnetic field ${\bf B}$ and the plasma velocity ${\bf u}$ 
are statistically similar in the different boxes, i.e., 
their spectra and their moments of the distribution function are the same in the different boxes. 
This essentially means that the statistics are scale-invariant in the inertial range.
The only difference is that the same fields are \emph{rescaled and mapped}, i.e., 
squeezed or expanded, into the new box with different dimensions.
Indeed, going from one box to another, the quantity that is actually changing is the size 
$L$ of the simulation box. Thus we can use the \emph{same realization} of the plasma velocity 
field of the RMHD simulation that is not dependent on the length $L$. 
Analogously the magnetic field, written in terms of the Alfv\'en velocity,
remains \emph{invariant} with changing the characteristic length $L$. 
A different discussion must be done for the Ohmic component of the 
electric field ${\bf E} = \eta{\bf j}$. 
From the dimensional analysis of this term, both the resistivity $[\eta]=[VL]$ and 
the current density $[j] = [B]/[L]$ are dependent on $L$. 
At smaller scales the current density ${\bf j}$ becomes stronger, while the resistivity decreases.
Thus we need to rescale them separately so that their product, i.e., the Ohmic electric field,
remains invariant
(assumes the same range of values and has the same spectra) in the different simulations.
We proceed as follow. First we assume that in each box the grid length $\Delta x$ is of 
the same order of the Kolmogorov microscale or dissipation scale $l_d$, determined by the 
condition that the dissipation rate equals 
the non linear energy transfer rate, $\tau_\textrm{diss}(l_d)  = \tau_\textrm{NL}(l_d)$ \citep{Frisch}.
For a given box, with characteristic length $L$ and velocity $u$,
this condition reads $l_d/\eta^2 = L/u(l_d/L)^{2/3}$, so at the length $l_d$ the resistivity is 
given by $\eta = (l_d^{4/3}/L^{1/3})u$, 
and $\eta$ decreases with increasing $L$.
The current $j$ evaluated at the length $l_d$, where the strongest current sheets are 
present, is of order $j = B/l_d \sim B/\Delta x$.
Considering the biggest and the smallest box in the models of grid size 
$\Delta x_B$ and $\Delta x_S$ respectively
(the subscripts $B$ and $S$ are associated with the biggest and the smallest box),
the condition $\Delta x_B \gg \Delta x_S$ ensures that $j_S \gg j_B$. Thus at smaller 
scales the current is stronger, as it is expected to be.
The requirement that the Ohmic electric field ${\bf E} = \eta {\bf j}$ spans the same 
range of values in the different boxes, i.e., $E_S = E_B$, leads to the condition 
\begin{equation}
 \left( \frac{l_d}{L} \right)  \sim \left( \frac{\Delta x}{L} \right) = \textrm{const}.
\label{eq:E_scaling}
\end{equation}
This is equivalent to assuming a constant magnetic Reynolds number $R_m=uL/\eta c^2$ in the different boxes.

As the parameter $\beta = \lambda_c/d_i$ in Equation~(\ref{eq:ODE2}) is the ratio between the correlation length 
and the ion inertial length (the largest and smallest physical scales in the system),
we ensure that the particle in each box has knowledge
of the real large-scale correlation length of the fields keeping
the value of parameter $\beta$ \emph{constant in all computational boxes}
and equal to its realistic value $5 \times 10^5$.
In all the simulations we maintain the same aspect ratio between the 
perpendicular and the parallel direction 
$l_\parallel=10\, l_\perp$ and periodic particle reinjection 
boundary conditions are used in all directions.  
One of the conditions respected by the multi-box method is the requirement that 
the particle gyroradii are not so large in the rescaled boxes that a particle can 
gyrate at a box size in the perpendicular directions.  
Thus avoidance of box-size  effects due to periodicity is highly desirable.  
The RMHD fields in the parallel direction are not periodic while the
particles that cross the $z$-boundary are re-injected at the other boundary, thus
they experience a small discontinuity in the physical fields.
However this approximation is not expected to have a signifcant 
impact on the statistics of the  
particle dynamics. In fact the current sheets are strongly elongated along $z$, and,
following the reduced MHD scaling, also the magnetic and velocity fields change
more slowly their values along this direction.
Thus the fields and structures experienced by the particle during its gyromotion do not change dramatically
during the crossing of the box boundary in the parallel direction.

The correlation length of photospheric convective motions is $\sim 1,000$\,km,
the biggest box in the perpendicular direction is extended for 4 correlation lenghts
$L_B = l_\perp = 4000$~km, with $256\times256$ grid points in the $x$-$y$ plane and 
$\Delta x = 3.9\times10^{-3} l_\perp \gg d_i$ (Figure~\ref{fig:box}). 
The perpendicular length of the smallest box measures 
$L_S = L_1 = 86d_i = 4.87\times10^{-5} l_\perp$,
with a grid of $256\times256$ in the $x$-$y$ plane and 
$\Delta x_1 = d_i/3 = 1.9\times10^{-7}  l_\perp$, smaller than
the minimum {\bf gyroradius}. 
Considering the initial condition for the particles, five boxes are needed 
for particles to travel the entire length of the coronal loop.

Ten thousand particles are injected in the smallest simulation box (hereafter BOX1) 
with random position and the same thermal velocity $v_\textrm{th} = 0.6\, v_{_{\!A}}$, 
corresponding to the coronal temperature $T=10^6$~K. 
In spherical coordinates, with the polar axis along the $z$-direction parallel to the mean 
magnetic field of strength $B_0$, particle velocity components are:
\begin{equation}
v_x = v\sin \theta \cos \phi \quad v_y = v\sin \theta \sin \phi  \quad v_z = v\cos \theta. 
\label{velocities}
\end{equation}
Particles initial velocities are randomly distributed in the gyrophase $\phi$ and pitch angle 
$\theta$.
Time is advanced with a fourth-order Runge-Kutta integration method with an adaptive 
time-step~\citep{pre92}. The values of the fields at each particle position are obtained by linear interpolation
in space from the grid of the RMHD simulation.
The trajectories are calculated using quasi-static approach in which magnetic and electric 
{\bf fields} do not change with time.

The simulation in BOX1 is stopped after  $5\times10^5 \tau_g = 0.15 t_{_{\!A}}$, 
where $ \tau_g = 2\pi/\Omega$ is the mean particle gyroperiod (computed in the mean field).
At this time we select the highest energy (H-E) particle population, 
about $\sim 10\%$ of the total BOX1 population, 
and inject them randomly in the next larger simulation box, BOX2.
The gyroradius of the particles that are transferred is $> 10^{-6}l_\perp$ 
at the end of the BOX1 simulation.
BOX2 has a grid scale of the order of their gyroradius ($\Delta x_2 = 10^{-6} l_\perp$) 
and a perpendicular length $L_2 = 2.56\times10^{-4}l_\perp$.
Analogously, the simulation in BOX2 is stopped at $t = t_{_{\!A}} \sim 3\times 10^6 \tau_g$ 
and a new high-energy population
(with $\rho \ge 10^{-5} l_\perp$)  is selected and randomly injected in the third box 
(BOX3), which has a grid scale 
$\Delta x_3 = 10^{-5} l_\perp$ and a perpendicular length $L_3 = 2.56\times10^{-3}l_\perp$. 
The BOX3 simulation is stopped at $t = 6 t_{_{\!A}} \sim 2\times 10^7 \tau_g$.
In each box the condition~(\ref{eq:E_scaling}) is satisfied.
The characteristic parameters used in the different simulations (box length $L$, grid size $\Delta x$, runtime $t_F$, average gyroradius $\bar{\rho}$)
are summarized in Table~\ref{tab1}.
The grid scale in the second box is about five times larger than in the first box, while maintaining the same number of grid points.  
Furthermore the grid scale in the third box is ten times larger than in the second box.
This essentially means that larger scale structures are present in larger boxes. These structures are required for particles with larger gyroradius 
to have a natural interaction with the fluctuations, including resonances.

At the end of the third simulations particles have reached energies 
of $0.1$ GeV and a non-relativistic approach
could be not suitable for higher energy. Moreover the absence in our model of an escape mechanism 
and the lack of self-consistency intrinsic in test-particle simulations could become an important limitation too, especially in presence of very high energy particles.
Thus, we decided to stop this study with the simulation performed in BOX3 (please also note that $L_B \neq L_3$).
\begin{table}
\center
\begin{tabular}{ | l | c | c | c | c |}
\hline
BOX           &  $L[ l_\perp]$                               &    $\Delta x[ l_\perp]$             &       $t_F[ t_{_{\!A}}]$    &    $\bar{\rho}[ l_\perp]$       \\            
\hline
$1$            &   {\bf $L_1 = 4.87\times10^{-5}$}         &       $1.9\times10^{-7}$          &        $0.15$                  &      $10^{-7}$\\ 
$2$            &    $L_2 =  2.56\times10^{-4}$      &       $10^{-6}$                         &        $1$                       &      $10^{-6}$ \\ 
$3$            &    $L_3 = 2.56\times10^{-3}$       &       $10^{-5}$                         &        $6$                       &      $10^{-5}$ \\ 
\hline
\end{tabular}
\caption{Values of the parameters used in the three simulations: box orthogonal length $L$, 
grid size $\Delta x = L/256$, runtime $t_F$ and average gyroradius $\bar{\rho}$, in units of the 
perpendicular cross section length scale of the physical loop  
$l_{\perp} = 4,000$\,km and Alfv\'en crossing time~$t_A=l_{\parallel}/c_A$ 
($l_\parallel = 10 l_\perp$). 
}
\label{tab1}
\end{table}

In the next section we show for each box the results corresponding to the total particle distribution
and separately for the associated high-energy (H-E) population. 

\section{Results} \label{sec:4}

\begin{figure}
\begin{center}
\includegraphics[width=8.5cm]{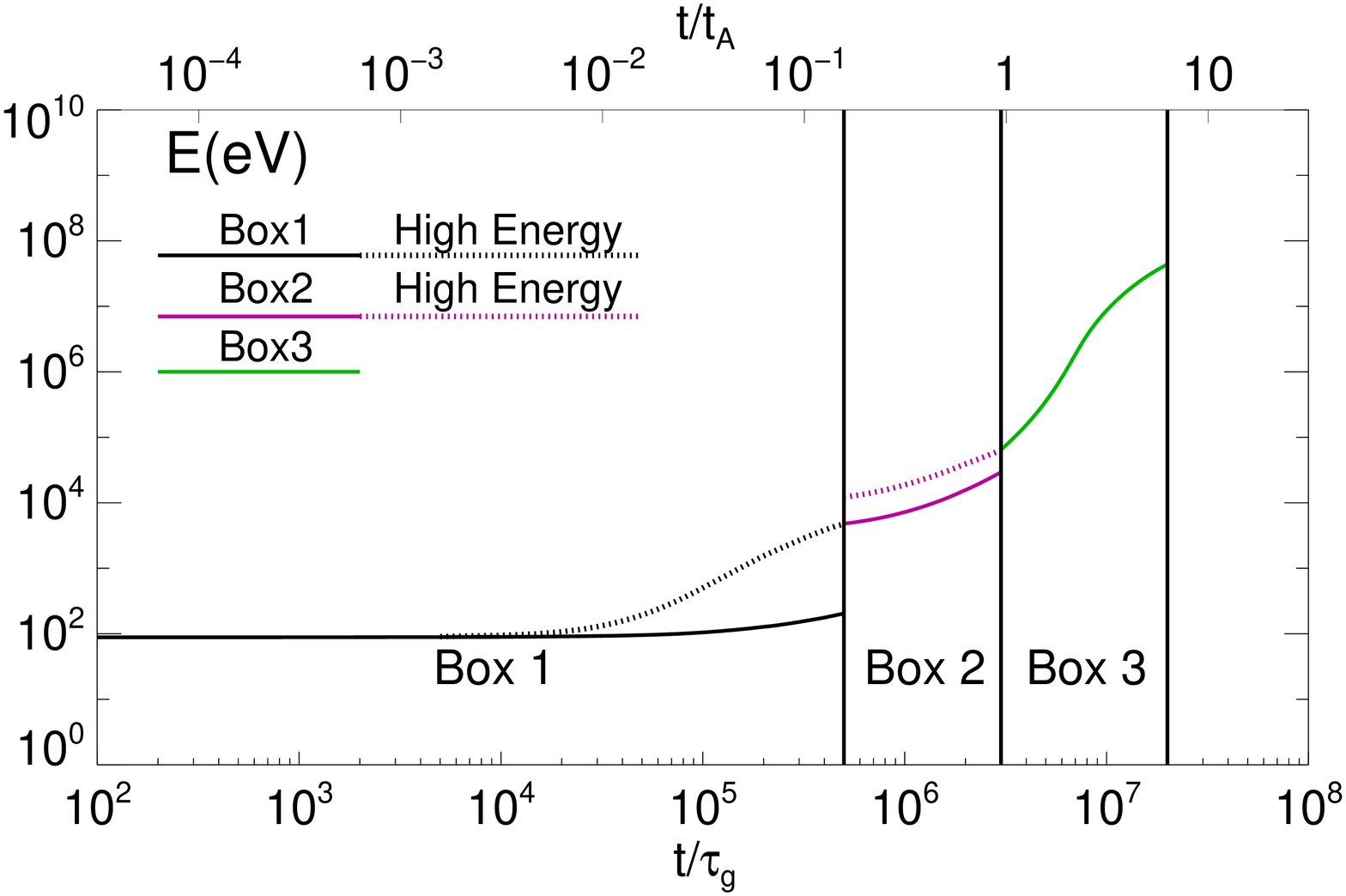}
\end{center}
\caption{Time history of the averaged energy E(eV) in the first (black line), in the second (purple line) and in the third (green line) box.
Continuous lines are for the total particle distribution in each box; dotted lines are for the selected high-energy population.}
\label{fig:energy}
\end{figure}

We begin by showing diagnostics that characterize the energization of the particle populations in time. 
Figure~\ref{fig:energy} shows the time history of the average particle energy $E$ in eV,
for the entire simulation spanning BOX1, BOX2 and BOX3. 
\begin{figure}
\begin{center}
\includegraphics[width=8.5cm]{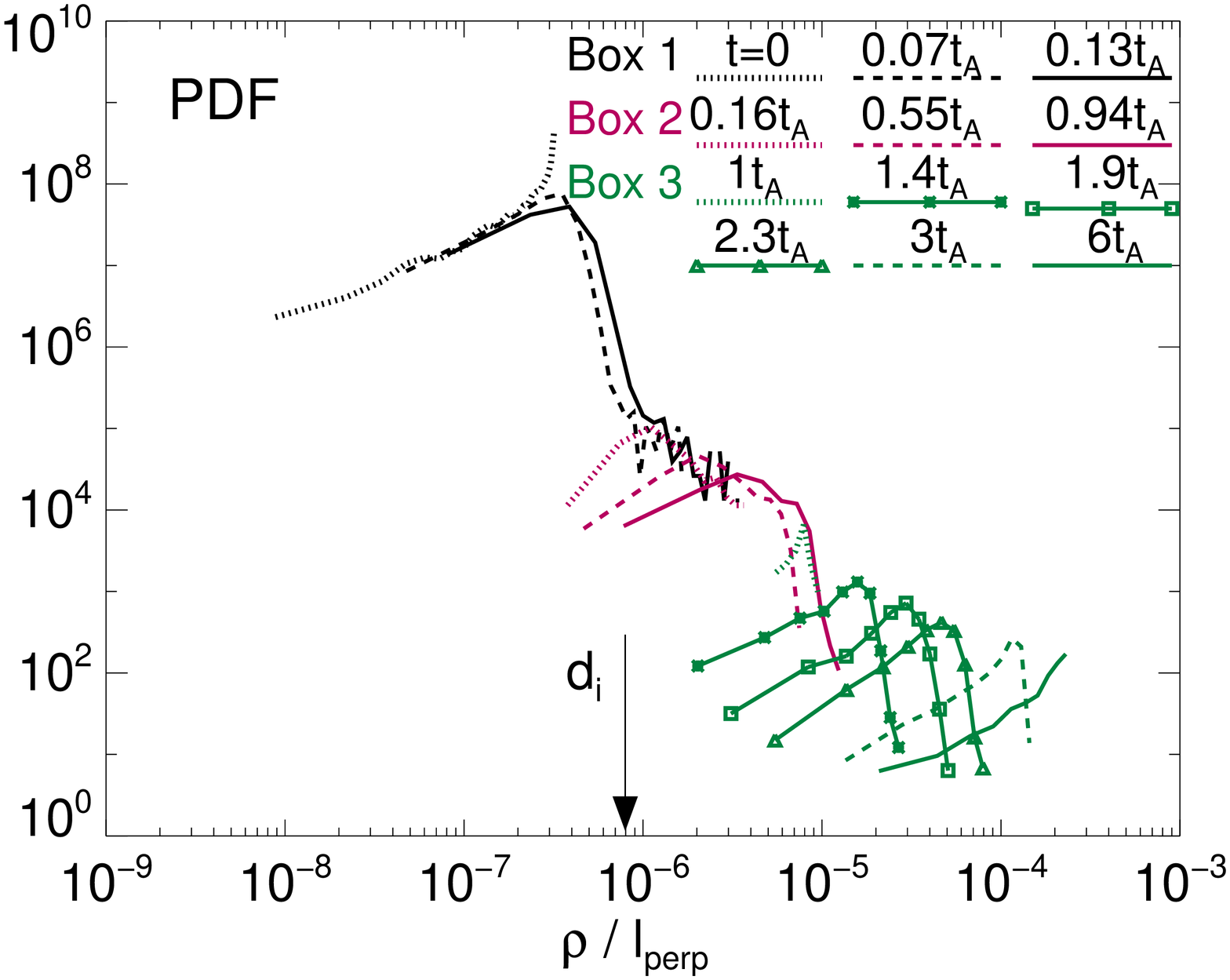}
\end{center}
\caption{Probability distribution functions (PDFs) of protons gyroradii $\rho$ in the first (black line), second (purple line), and in third (green line) box.
Different lines correspond to different times in the evolution, as shown in the legend.}
\label{fig:rho}
\end{figure}
Figure~\ref{fig:rho} shows the probability distribution functions (PDFs) of proton gyroradius, 
$\rho = v_\perp^2/|{\bf B}|$, at different stages of the evolution. 
The PDF in BOX2 and in BOX3 are normalized to $1/N_1^*$ and $1/N_2^*$, 
where $N_1^*$ and $N_2^*$ are the percentage of particles selected at the end of the previous simulation and 
re-injected randomly in BOX2 and BOX3, respectively.
We want to caution the reader at this point 
that while we present results in physical units, our goal is to study the
physical mechanisms and not to reproduce any real datasets. 
There are numerous reasons that our simulation data would be expected 
to differ from space and astrophysical data, for example, 
the lack of consideration of mechanisms of escape, and the lack 
of self-consistency (test-particles), as well as 
other features dictated by the numerics.

The initial energy distribution is sharply peaked (delta function) at $89$eV. 
At the end of the simulation in BOX1, after $0.13 t_{_{\!A}}$, the energy of the H-E particles (dashed black line) 
is more than one order of magnitude larger than the energy of the total population.
When simulation starts, ten thousand particles are injected in the box; when it is stopped only 
$10\%$ of the overall population gain sufficient energy for their gyroradii to be an order of magnitude bigger than
the grid scale in BOX1, and therefore comparable with the grid size in BOX2,
so that the particle can sample the structure of the fields in the second box. 
Indeed the H-E particles, increasing their energy, gyrate with a larger {\bf gyroradius} and are allowed 
to sample and interact with different regions of the field.
In contrast, particles that do not gain a sufficiently large amount of energy continue 
to gyrate with a gyroradius that is not large enough to explore different field structures, 
as if they were ``confined'' to the same range of scales, and to the same region of space. 

\begin{figure}
\begin{center}
\includegraphics[width=4.2cm]{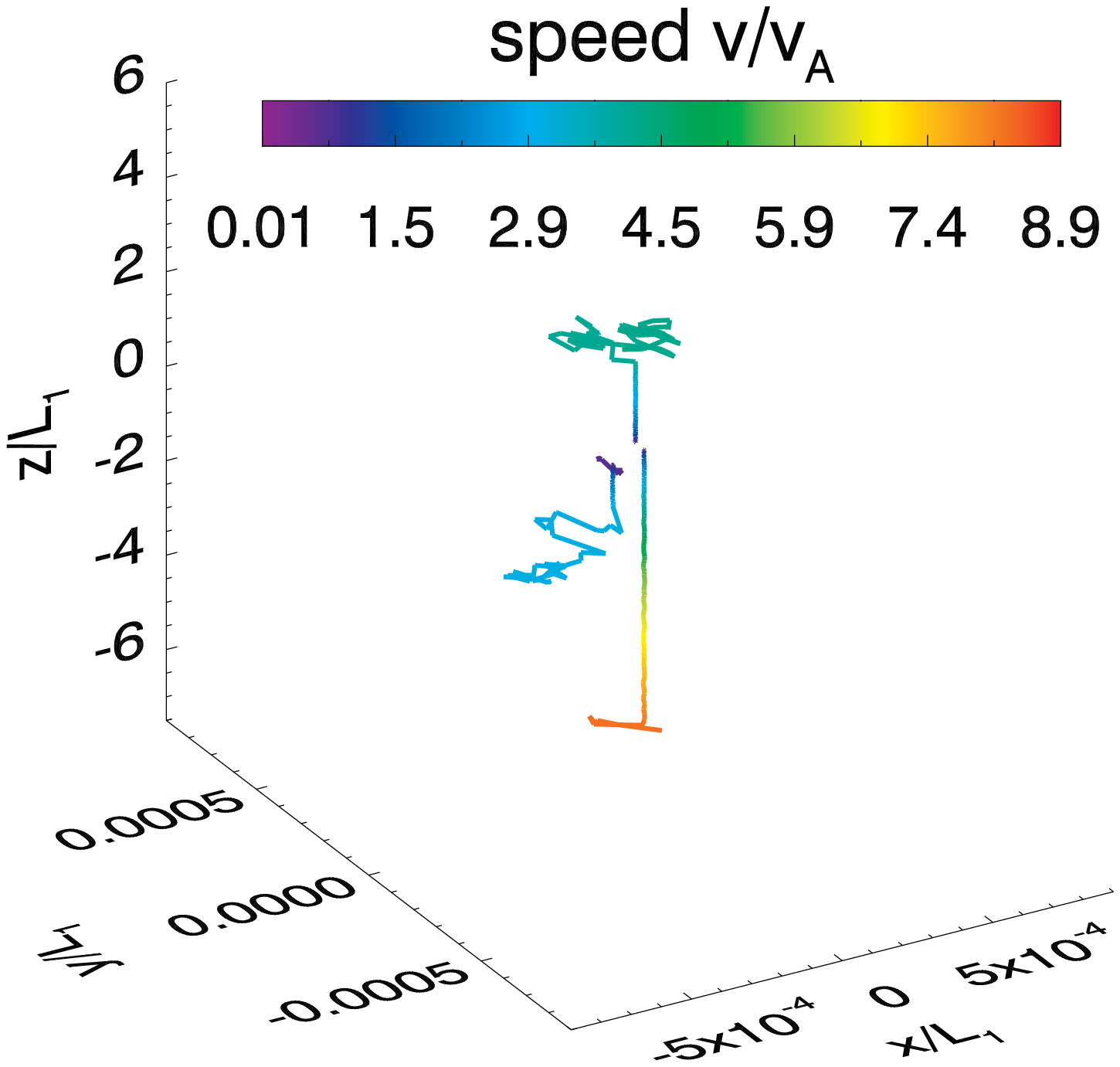}
\includegraphics[width=4.2cm]{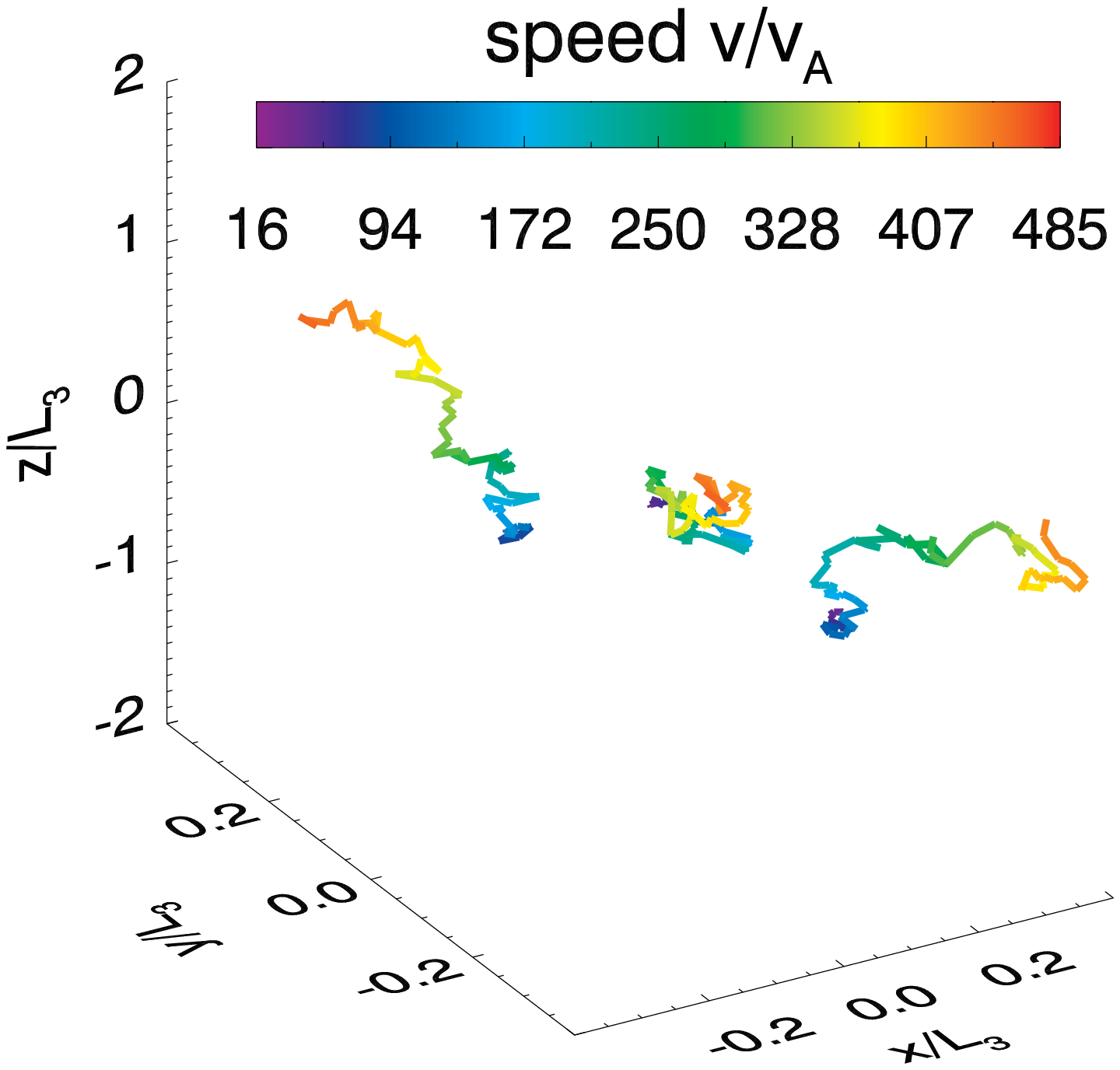}
\end{center}
\caption{Trajectories of three of the most energetic particles in BOX1 (on the left) and in BOX3 (on the right).
The colors in the trajectories indicate the speed of the particle.}
\label{fig:traj}
\end{figure}

This idea is reinforced examining the trajectories performed by the particles in BOX1, shown in the plot on the left of Figure~\ref{fig:traj}. 
Three different trajectories are highlighted in the picture, 
with changing colors corresponding to increasing protons speed.
Particles are observed to accelerate most when their trajectories
are aligned to the mean axial field ${\bf B}_0$.
For instance, two of them gain energy traveling a short distance in the axial 
direction and then start to rattle around trapped at constant $z$.
On the other hand, the third particle with the trajectory more elongated along $z$ rapidly ascends gaining energy. 
This elevator-like path is associated with sheet like structures of electric current 
density that are extended in the axial direction, and provide channels for particle motion.  
Because at this early stage the  gyroradius is comparable or smaller than the ion inertial scale $d_i$
($\rho_\textrm{max}=0.6d_i$ at the injection, see black dotted line in Figure~\ref{fig:rho}), 
when a proton encounters a current channel, it is likely to remain inside the channel over several gyroperiods.

In this scenario, protons that find strong currents will reach larger energy mostly 
in the parallel direction, while protons that encounter only average or 
small currents will only acquire a moderate energy. 
It is also important to notice at this point that in BOX1 the largest difference between the averaged 
energy of the total distribution and of the H-E population is observed. 
This suggests that, although the energies reached at the end of the first
stage are not extremely high, the H-E population is selected at the beginning of the evolution, 
namely for $t \ll t_{_{\!A}}$, by the effect of the Ohmic electric field acting in the direction parallel to the guide field 
${\bf B}_0$, along which the current density structures are preferentially aligned.
Note that near-alignment of parallel current channels with a strong imposed magnetic field
is expected for MHD flows that are mainly incompressive in nature \cite[see, e.g.,][]{OughtonEA94}.

\begin{figure}
\begin{center}
\includegraphics[width=8.5cm]{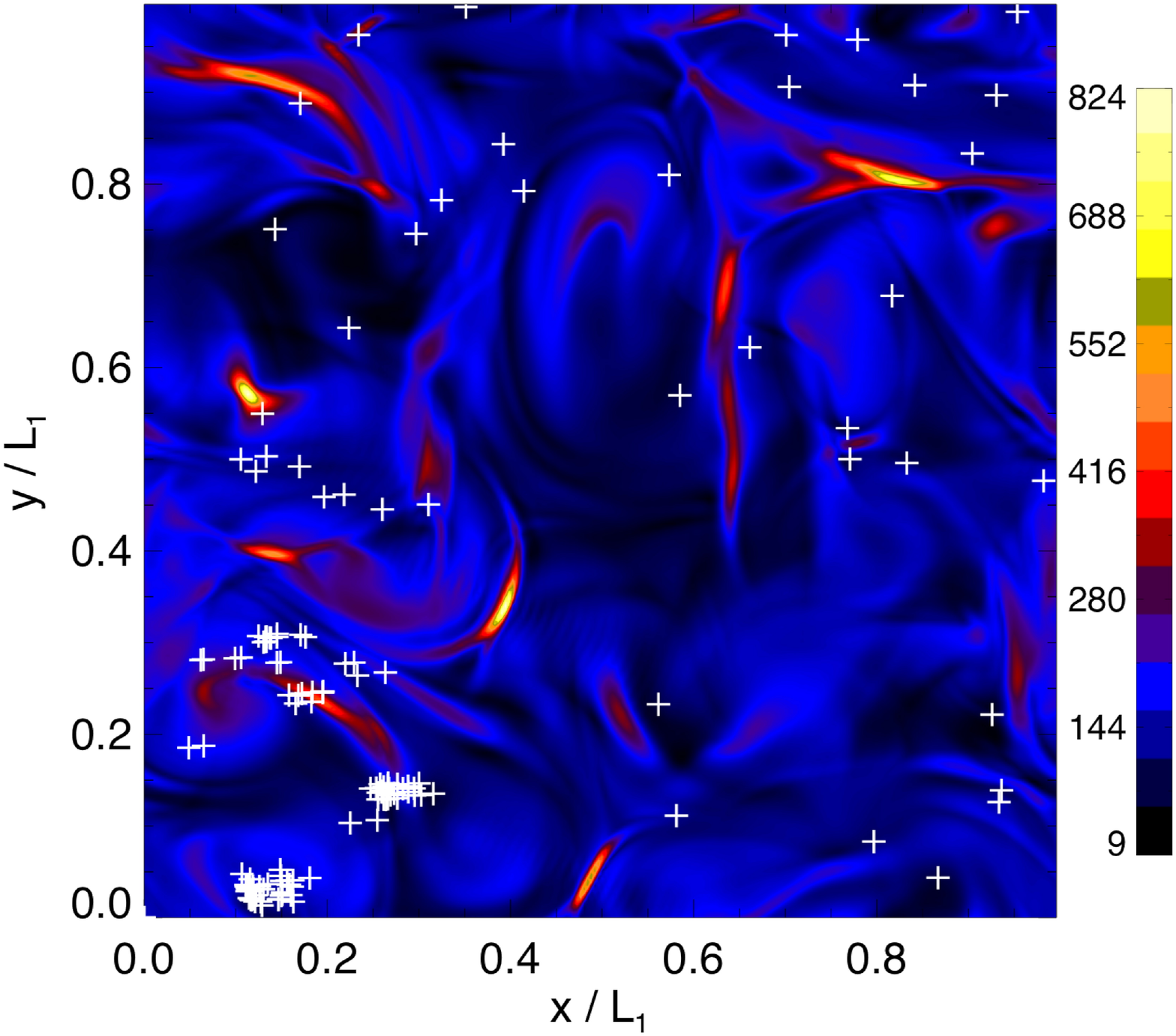}
\includegraphics[width=8.5cm]{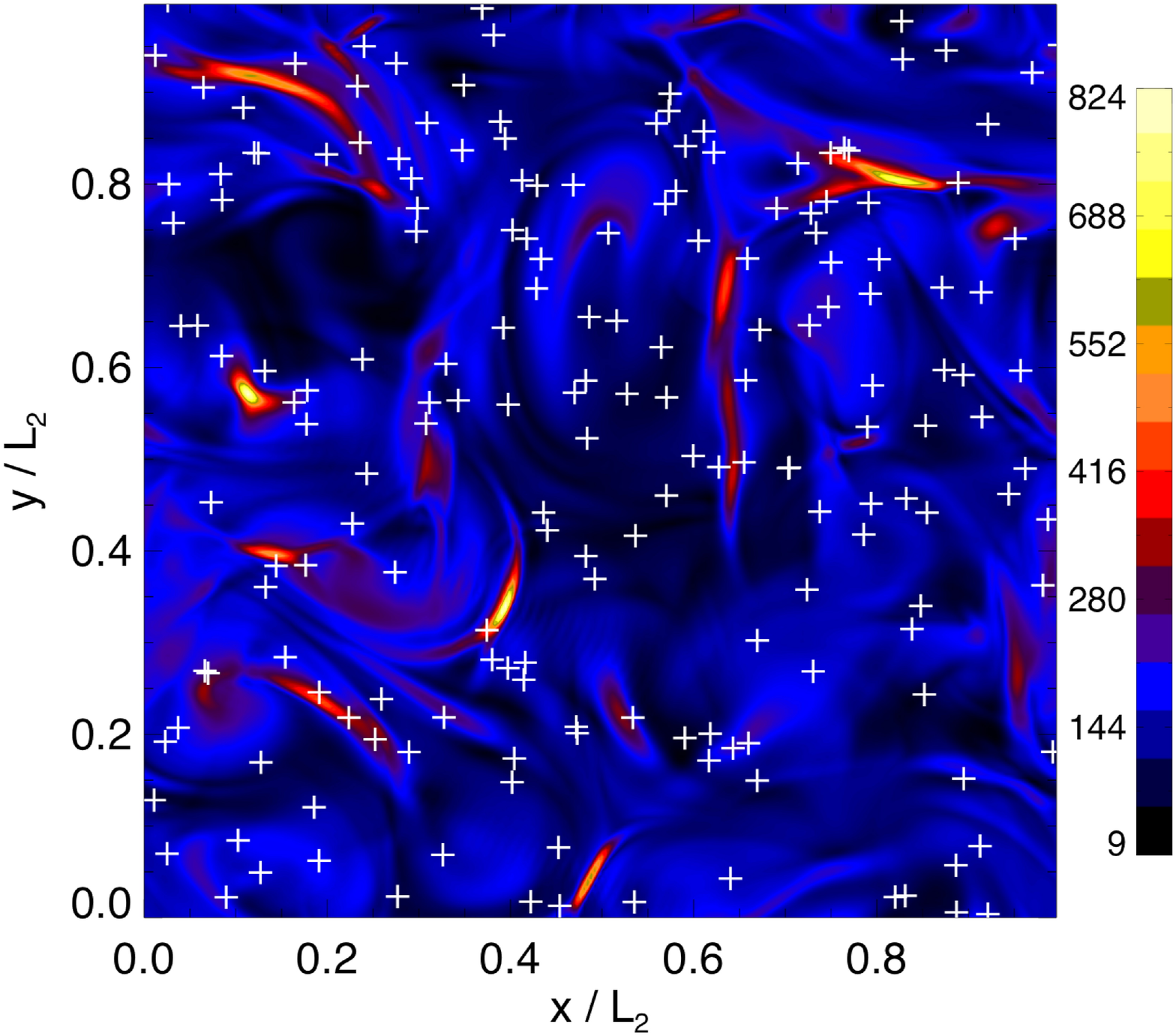}
\end{center}
\caption{Contour plot of the absolute value of the current averaged in $z$, $\langle | j | \rangle_z$. 
The white crosses represent the position of the high-energy particles in BOX1, top panel, and in 
BOX2, bottom panel.}
\label{fig:jmap}
\end{figure}
\begin{figure*}
\begin{center}
\includegraphics[width=18.cm]{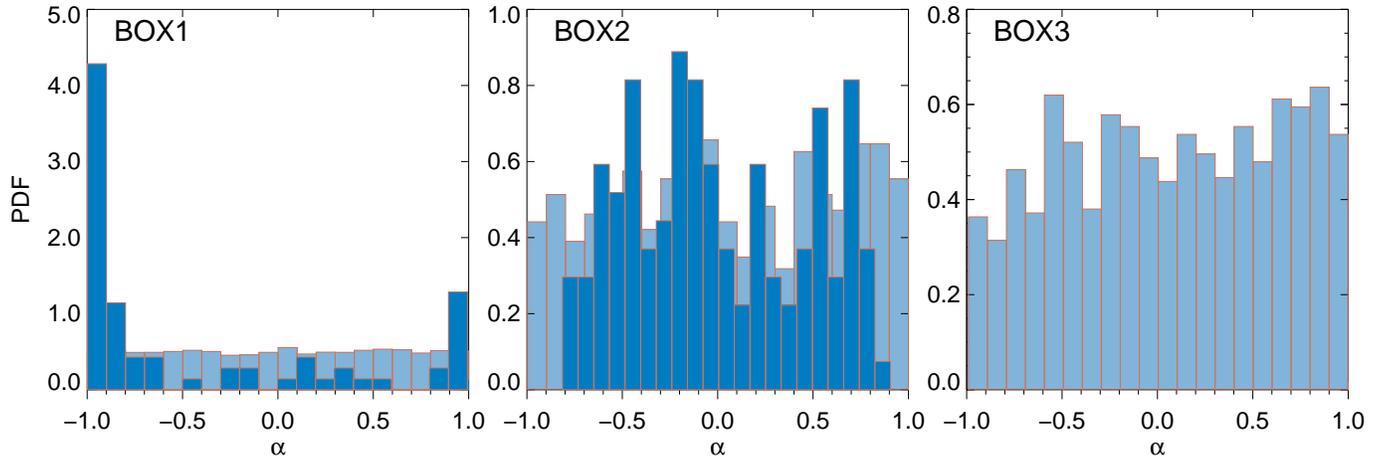}
\end{center}
\caption{PDF of particles pitch angle $\alpha$ in BOX1, left plot, BOX2, central plot, and BOX3, right plot, 
for the total population (light blue) and the high-energy population, dark blue.}
\label{fig:pitch}
\end{figure*}

In the second box, we inject $1000$ particles and follow their dynamics. 
These particles are taken from the high energy tail of the distribution at the end of the simulation in BOX1, 
as shown by the overlapping of the gyroradius PDFs at the beginning of the simulation in BOX 2 (Figure~\ref{fig:rho}).
The energy difference between the H-E population and the total distribution (Figure~\ref{fig:energy}) is substantially reduced,
reinforcing the idea that the effects acting at the smallest scale are responsible for the selection of the high-energy particles.

The analysis of the particle trajectories in BOX2 (not shown) in comparison to 
BOX1 indicates that much fewer particles travel almost undisturbed along the axial direction.
Indeed, if at the injection few particles gyrate with a gyroradius smaller or comparable with the 
current sheet thickness, then at later times (Figure~\ref{fig:rho}) the probability 
distribution function of $\rho$ moves toward larger values, i.e., $\rho \gg d_i$. 
As a consequence, protons do not remain in the current channels and they do not see a coherent 
parallel current, which could produce a net energization effect in the parallel direction, 
as happened previously in BOX1. Nonetheless, a net increase of the average energy is observed.

The same considerations enter the analysis of the dynamics in BOX3.  
It appears that all the particles undergo the same acceleration mechanism
here as in BOX2, reaching almost $0.1$GeV (green line in Figure~\ref{fig:energy}).
The motion in BOX3, shown in Figure~\ref{fig:traj}, 
cannot be associated with the presence of strong current channels.
Indeed, at the end of the third simulation all the particles have a gyroradius 
$\rho \sim 10^{-4}\, l_\perp$ (Figure~\ref{fig:rho}),
greater than the thickness of the strong current  sheets. 
Moreover it is not possible to differentiate between two separate populations of particles.

A closer look to the H-E particles is required for a better understanding of their dynamics. 
In Figure~\ref{fig:jmap} we show the position of the H-E particles in BOX1 and in BOX2, 
together with contour plots of the absolute value of the current averaged in $z$, $\langle | j | \rangle_z$.
This picture shows qualitatively that two different acceleration mechanisms are acting 
during different times of the particles evolution.
While the positions of the H-E particles are more or less equally distributed within the simulation BOX2, 
(with however some ``voids'' in evidence), the clusters of particles close to
the current sheets structures in BOX1 indicates energization
due to the Ohmic electric field within the $z$-aligned current channels.
Eventually some particles pitch angle scatter out of the channel, or, 
after attaining higher energy, leave the current channel because their gyroradius increases.
This explains the presence of energetic particles in BOX1 further from the strongest current sheets.

Another way to study particle motions is to examine pitch angle distributions. 
In Figure~\ref{fig:pitch} the PDFs of the cosine of pitch angle 
$\alpha = v_\parallel/|v|$ in the three different boxes are shown. 
In BOX1 and in BOX2, distributions of $\alpha$ are shown for both the 
H-E population (dark blue) and the total population (light blue). 
As pointed before, in the third box it is not possible to pick out a 
high-energy distribution because all the particles reach 
more or less the same final energy. 
Thus, the PDF in BOX3 (right plot) has been computed considering all the particles in the simulation.
The H-E particles in BOX1 are mainly aligned to the guide field, $\alpha = \pm 1$, 
reflecting the fact that the accelerating field is in the $z$ direction. 
However in BOX2 the PDF of pitch angle of the H-E distribution moves away from $\alpha = \pm 1$ 
and is peaked around  $\alpha = 0$, indicating that the acceleration 
process is taking place mainly in the direction perpendicular to ${\bf B}_0$.
The pitch angle distributions support the picture of a 
two-stage acceleration process, which we further examine below. 

\subsection{On the nature of the perpendicular electric field}

We now  analyze the different terms in the equation of motion to explain qualitatively 
the observed dynamics of the protons. In this analysis we follow the procedure described in \cite{dmt04}.
For a general MHD representation of the plasma flows and fields, the equation for the parallel 
($\mathbf{v}_\parallel$) and the transverse ($\mathbf{v}_\perp$) proton velocities are given by:
{\small
\begin{equation}\label{eq:par}
\frac{d{\bf v}_\parallel}{dt} = \beta 
\left[ ({\bf v}_\perp - {\bf u}_\perp) \times {\bf b}_\perp  + \eta {\bf j}_\parallel \right],
\end{equation}
\begin{equation}\label{eq:perp}
\frac{d{\bf v}_\perp}{dt} = \beta 
\left[ ({\bf v}_\perp - {\bf u}_\perp) \times {\bf B}_0 + ({\bf v}_\parallel - {\bf u}_\parallel) 
\times {\bf b}_\perp + \eta {\bf j}_\perp \right],\\[.1em]
\end{equation}
}%
where ${\bf u}_\perp$ and ${\bf b}_\perp$ are the components of the plasma velocity and magnetic field 
perpendicular to the mean field ${\bf B}_0$, and ${\bf u}_\parallel$ is the parallel plasma velocity.
If we specialize these equations to RMHD fields, such as those used in the present study, 
several terms in Equations~(\ref{eq:par}) and (\ref{eq:perp}) vanish identically because 
${\bf u}_\parallel \equiv 0$ and ${\bf j}_\perp \equiv 0$ in RMHD. 
Moreover, given the ordering of RMHD fields ${\bf b}_\perp$ is small ($b_\perp < B_0$).

The terms in Equation~(\ref{eq:par}) all remain in RMHD, but in this highly anisotropic  
turbulence regime the inductive term, ${\bf u}_\perp \times {\bf b}_\perp$, will be small 
compared to $\eta {\bf j}_\parallel$ near the current sheets, and  it will be negligible elsewhere
as in the low corona velocity fluctuations are concentrated around current sheets \citep{rv11}. 
Thus the Ohmic electric field is the dominant term in the equation for the parallel velocity, 
and its efficiency is predominant at smaller scales, as we showed above.
\begin{figure*}
\begin{center}
\includegraphics[width=18.cm]{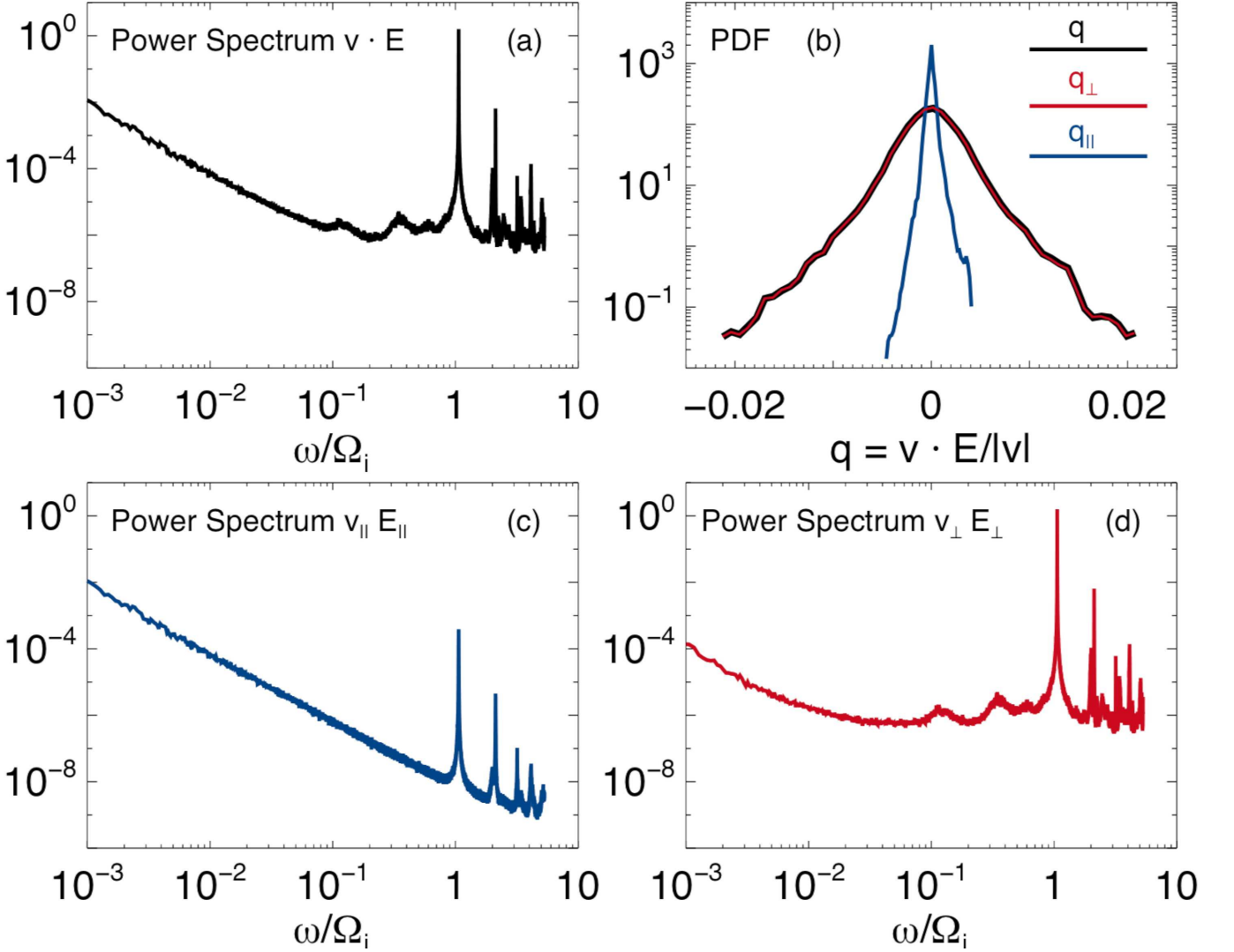}
\end{center}
\caption{Power spectrum of the work done by the electric field on the particle 
$({\bf v} \cdot {\bf E})$ (panel~a). 
The parallel (panel~c) and the perpendicular (panel~d) terms are shown separately. 
Panel~b shows the PDF of the quantity $q = ({\bf v} \cdot {\bf E})/ |v|$ (black line).
The PDFs of the parallel $q_{\parallel} = (v_\parallel E_\parallel)/|v|$ 
and perpendicular $q_{\perp}  = (v_\perp E_\perp)/|v|$ terms are represented with blue and red 
lines, respectively. Both spectra and the PDFs are computed from particle trajectories in Box 3. }
\label{fig:res}
\end{figure*}

After setting $\mathbf{u}_\parallel = \mathbf{j}_\perp \equiv 0$ in Equation~(\ref{eq:perp}),
of the remaining terms it is the first on the right-hand 
side that can effectively accelerate particles. 
It gives rise to a drift motion with velocity 
${\bf u}_\perp$ plus a gyromotion around the magnetic field ${\bf B}_0$.
For particles temporarily confined within a current channel, the parallel Ohmic field $\eta \mathbf{j}_\parallel$ will be dominant. 
However, as soon as their gyroradii become comparable with the current sheet thickness $d_i$, protons will see a transverse 
MHD-induced electric field ${\bf E}_\perp = -({\bf u}_\perp \times B_0{\bf e}_z)/c$,
that varies on scales comparable to their gyroradius. 
If kicks of this electric field are in phase with the proton velocity, this results in a net increase of the perpendicular energy.
This may be viewed as a betatron-like resonance process \citep[for a description of the betatron mechanism see][]{Swann33}.
Thus, variations of the transverse electric field are produced by variations in the plasma velocity ${\bf u}_\perp$.
The plasma velocity is concentrated around the current sheets, that are indeed embedded in a vorticity 
quadruple. Therefore particle acceleration associated with the perpendicular electric field component will act more strongly
in the region near current sheets, including the interior of flux tubes bordering the current sheets.

In order to verify the effectiveness of this mechanism, we computed the work done by the 
electric field on particles moving with velocity ${\bf v}$, namely $({\bf v} \cdot {\bf E})$. 
This calculation is done in BOX3, where the betatron-resonant process is thought to dominate.
Figure~\ref{fig:res}a shows the power spectrum of $({\bf v} \cdot {\bf E})$ averaged over $100$ particles. 
The parallel and the perpendicular terms in the product,  
$v_\parallel E_\parallel$ and $v_\perp E_\perp$, are both shown, and 
the frequencies are expressed in term of the characteristic ion gyrofrequency $\Omega_i$. 
For this analysis, particle trajectories have been integrated over $10^4$ gyroperiods,
saving the required data every $\Delta t = 0.1 \tau_g$.
It is clear that the process involves a substantial degree of resonance, as strong peaks are 
found for $\omega = \Omega_i$ and for the successive harmonics, from $n=2$ to  $n=5$.
These feature indicates that a strong coupling between the plasma and the particles is occurring at the particle gyrofrequency.

To verify that protons have been effectively accelerated in the direction perpendicular to ${\bf B}_0$, we computed
the PDF of $q = ({\bf v} \cdot {\bf E})/|v|$ together with the PDFs of the parallel $q_\parallel = (v_\parallel E_\parallel)/|v|$ 
and perpendicular $q_\perp = (v_\perp E_\perp)/|v|$ terms. These quantities are shown in Figure~\ref{fig:res}b. 
The PDFs have been computed considering all the values along the trajectories of $100$ particles.
The PDF of the total $q$ and the PDF of the perpendicular term $q_\perp$ are nearly indistinguishable. 
The PDF of $q_\parallel$ is narrower, showing that the energization is larger in the perpendicular direction. 
As expected $q$, $q_\parallel$ and $q_\perp$ display both positive and negative values. 
Indeed, throughout their evolution, particles can be accelerated but they can slow down as well. 
However the average values of the three distribution are positive, confirming that a net acceleration is occurring.
The average values and their standard deviations are
${\bar q} = 1.2 \times 10^{-6}$ and $\sigma = 2.7 \times 10^{-3}$,  
${\bar q_\parallel} = 8.3 \times 10^{-7}$ and $\sigma_\parallel = 3.8 \times 10^{-4}$, 
${\bar q_\perp} = 4.7 \times 10^{-7}$  and $\sigma_\perp = 2.7 \times 10^{-3}$.

Moreover in Fig 7, the lower frequency part of the total power spectrum (panel a), 
i.e., for frequencies lower than $0.1\Omega_i$, is dominated by the parallel terms, 
while for higher frequencies the perpendicular component dictates the shape of the spectrum. 
%
       The $\sim 1/\omega^2$ behavior of the parallel frequency spectrum suggests second order 
       Fermi acceleration of a standard type, in which scattering from counter-propagating fluctuations
       causes a momentum diffusion. However it is also clear that this is not the dominant acceleration
       mechanism at this stage. Perpendicular acceleration accounts for about two orders 
       of magnitude more power (panel d) and consists of systematically greater magnitudes of acceleration 
       (panel b). We may then conclude that the system remains complex, with several mechanisms
       involved at this stage, but according to several measures shown here, the betatron 
       mechanism may be viewed as the strongest mechanism that we have identified. It is also 
       clear that not only a separation of length scales is involved, but also a separation of 
       times scales .

As we pointed out above, the energy increase in the perpendicular direction arises if variations of the plasma
transverse velocity field ${\bf u}_\perp$ are in phase with the gyromotion of the particle.
From this perspective the process can be thought as resonant.
However, it is important to emphasize that the mechanism described should be considered as a {\it first-order} process, 
being caused by a convective electric field ${\bf E}_\perp = -({\bf u}_\perp \times B_0{\bf e}_z)/c$.
It is also important to point that, although it represents the main acceleration mechanism in the second and in the third box,
the coupling between the convective electric field and the particles can take place 
as soon as the particle gyroradius becomes bigger than the ion inertial scale $d_i$.

\begin{figure}
\begin{center}
\includegraphics[width=8.5cm]{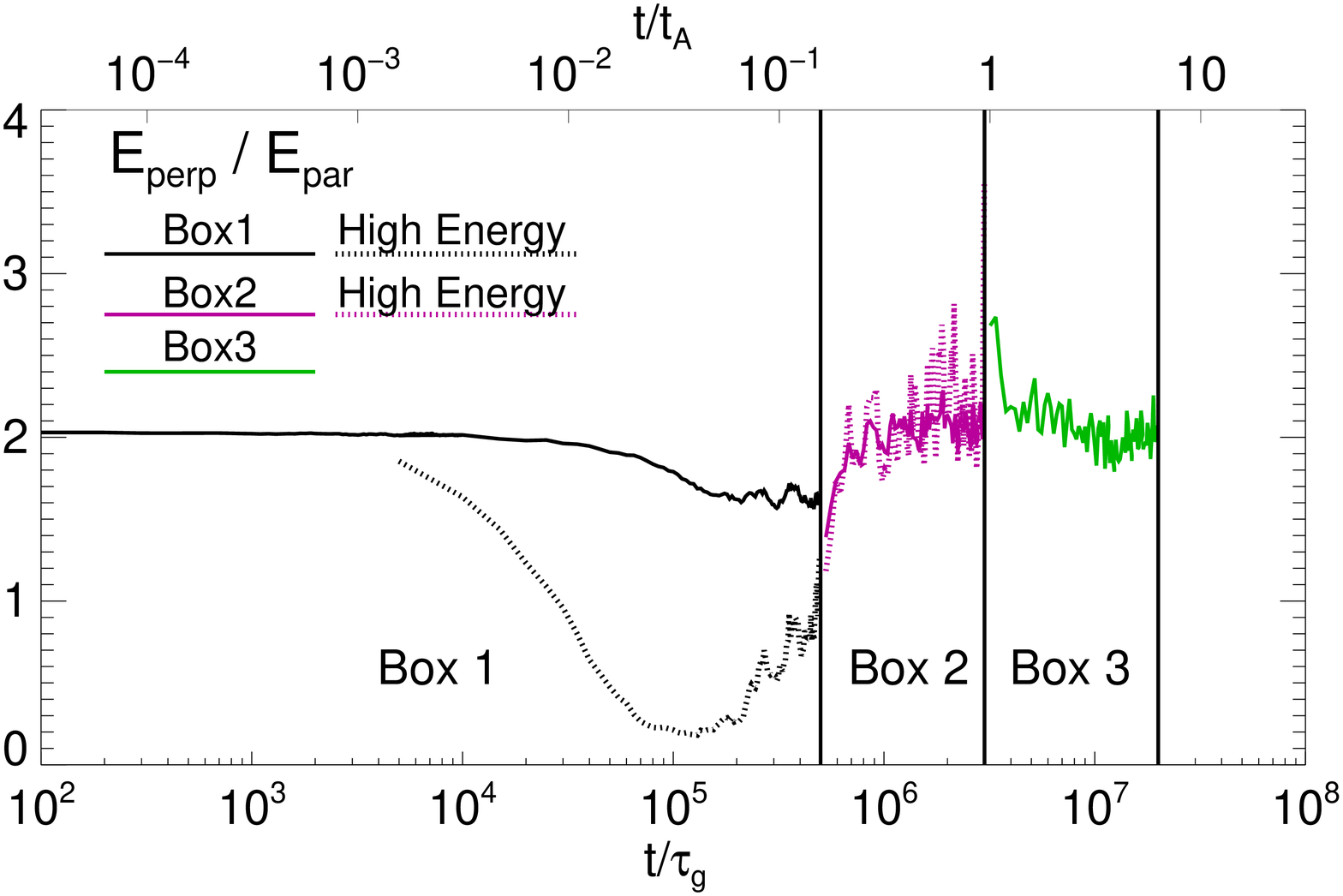}
\end{center}
\caption{Time history of the fraction between the perpendicular and the parallel energy, $E_\perp/E_\parallel$, in the first (black line), 
in the second (purple line) and in the third (green line) box. 
Continuous and dotted lines represent the total particle distribution and the selected high-energy population, respectively.}
\label{fig:Eratio}
\end{figure}

More investigations have been conducted to characterize better the energization that occurs during the different stages.
Figure~\ref{fig:Eratio} shows the time history of the ratio of average perpendicular energy and 
average parallel energy $E_\perp / E_\parallel$ in the different boxes for the total population and for the H-E particles.
It is clear from the evolution of $E_\perp / E_\parallel$ in BOX1 that the H-E particles are preferentially accelerated
in the direction parallel to ${\bf B}_0$.
When particle energy is sufficiently high, pitch angle scattering can transfer momentum from the parallel to the perpendicular
direction, thus explaining the increase of the ratio at the end of the BOX1 simulation.
Subsequently, the ratio $E_\perp/E_\parallel$ rapidly attains values near $2$, i.e., 
the distribution tends to isotropize because of pitch angle scattering, as shown above 
by the PDFs of $\alpha$ for the H-E population in BOX3 and in BOX2. 
However in BOX2 it is still possible to observe strong peaks of $E_\perp/E_\parallel > 2$, 
testifying a strong acceleration in the perpendicular direction. 

\section{Conclusions} \label{sec:5}

In this paper we investigated particle acceleration in turbulent electromagnetic fields
embedded in a strong guide magnetic field,
an important issue in astrophysical, space and laboratory plasmas.
We performed test particle simulations for protons in turbulent fields obtained 
from direct numerical solutions of reduced Magnetohydrodynamic turbulence
(for a detailed description of the RMHD simulation see \cite{rved08}).
Although the test particle approach is not self-consistent, it provides a useful bridge between 
the large-scale macroscopic MHD and a more realistic particle kinetic physics description.
The geometry of the system and the typical parameters used in the simulations
are well-suited for the description of particle acceleration in coronal loops. 

A common problem in the numerical description of plasma and/or particle 
dynamics in space plasmas is the involvement of a huge range of length and time scales.
In the case of coronal loops the large scale motion occurs on a distance 
comparable with the loop perpendicular length (thousands kilometers), but the acceleration process
takes place at different stages and/or locations during the evolution of the system
and may require the participation of several mechanisms.
Therefore it can be considered as a multi-stage process, and the numerical study
may require that the various length and time scales should be followed with different
time and spatial resolutions. In the study of this kind of system, the acceleration may 
be associated with strong and sudden changes, including changes of  
magnetic topology near reconnection sites as well as strong discontinuities.
Consequently the use of the guiding center equations may be inappropriate, 
because non-adiabatic behavior may develop on local scales.

Motivated by these reasons, and to maintain the possibility 
of gyro-resonant interactions at all phases of the acceleration process, 
we have developed a {\it multi-box, multi-scale} technique to solve the nonrelativistic equation
of motion of a particle moving by the action of electromagnetic fields (see Section~\ref{sec:3}).
This technique allowed us to resolve 
an extended range of scales present in the system, namely from the 
ion inertial scale $d_i$, of the order of $1$\,m, up to macroscopic scales
of the order of $10$\,km ($1/100$th of the correlation scale, the 
photospheric convection scale $\sim 10^3$\,km), 
resolving at the same time the particle gyroradius.
An important feature of the multi-box method is that the scaling of the MHD fields is
controlled so that the magnetic field, plasma velocity, and electric field take on similar values in the rescaled 
boxes as their size varies from the initial smallest box to the largest final box. 
This amounts to changing length scales and resistivities in the 
sequence of boxes to maintain the required statistical continuity.

Employing the multi-box strategy, we performed a numerical experiment that 
shows how acceleration takes place typically in a two stage process. 
First, a small fraction of initially loaded low energy particles are entrained 
in elongated current sheets long enough to gain substantial  
energy in response to the parallel electric field in the current sheet.
High parallel speeds are attained in this stage.
Gradually these particles escape the current channels due to pitch angle
scattering and due to their larger gyroradius after energization.   
In the second stage, particles encounter nonuniform 
in-plane electric fields that enable a resonant betatron acceleration for those
particles that remain entrained in these regions for a suitable amount of time.
In this case the acceleration is mainly perpendicular. However once again pitch 
angle scattering tend to isotropize the particles, which also then facilitates their 
escape from  the accelerating region. 

A point of interest is that the two stage acceleration process described here is fully consistent 
with important features of test particle acceleration in turbulence that have been reported previously.
For example, the initial parallel acceleration of test particles in current channels 
associated with two dimensional reconnection was reported some time ago by \cite{amb88}. 
There, too, it was observed that the energization is limited by the time of temporary trapping of particles near the acceleration region.
The role of pitch angle scattering in disrupting the parallel acceleration process was also described.
Furthermore it is likely that secondary islands near or in current sheets may enhance the ability of 
sheet-like structures to entrain and accelerate particles \citep{amb88,DrakeEA06}.
We have not attempted to describe the effect of secondary islands here.

It is also noteworthy that the parallel and perpendicular stages of acceleration that we described 
were anticipated, although as separate effects, in the study by \cite{dmt04}. 
In that case, using a single simulation box, it was shown that at early times, 
protons experience perpendicular acceleration, while electrons experience parallel acceleration.
The consistency of that picture with the present one is established by noting that in that study, 
particles were initialized with zero speed, but quickly acquire speeds comparable to the Alfv\'en speed. 
However the current sheets have thickness comparable to the ion inertial scale, 
so the protons almost immediately encounter the edges of the current sheets and therefore enter 
the perpendicular ``stage two'' process described here.
Given their much lower inertia, electrons have very small gyroradii when moving at the 
Alfv\'en speed, and so can experience what we presently call the stage one parallel acceleration process, 
gaining substantial energy in a short time. 
The multi-box method allows the computation to follow both stages one and two for a single particle type, 
while maintaining resonant couplings that are central to spatial transport and 
therefore contributing to entrainment and escape processes.

For the particular example appropriate to coronal loops that we showed here, by the end of the third box 
in the multi-box sequence particles attained energies up to $0.1$ Gev.
In order to cover a realistic large parallel coronal loop length, another two simulations 
in larger boxes would need to be performed.
However, at the end of the third simulations particles have reached energies of 0.1 GeV and a non-relativistic approach
could not be not suitable for higher energy.
Continuing in this way, higher energies are expected at the end of the largest scale simulation,
and  an ``escape'' mechanism mimicking the collision of particles with the denser
chromospheric layer \citep[e.g.,][]{GontiEA13} should be included. 
Furthermore it must be pointed out that the lack of self-consistency intrinsic in test-particle simulations
could become at some point an important limitation too, especially
for the highest energy particles.

Although a refinement of the {\it multi-box} technique may be also required, 
the results shown in this paper indicate a promising way of understanding energetic particles 
in astrophysical, space, and laboratory environments and studying some basic issues 
in the complex topic of bridging the MHD and kinetic descriptions of a plasma.

\acknowledgments

This research supported in part by NASA Heliophysics Theory program NNX11AJ44G, 
NSF Solar Terrestrial and SHINE programs AGS-1063439
\& AGS-1156094,  NASA Magnetosphere Multiscale mission through 
the Theory and Modeling Team, and by the Solar Probe Plus Project through the ISIS Theory team, 
and by EU Marie Curie project ``Turboplasmas'' at the University of Calabria.


\begin{thebibliography}{}
%
\bibitem[Ambrosiano et al.(1988)]{amb88} 
Ambrosiano, J.~J., Matthaeus, W.~H., Goldstein, M.~L., \& Plante, D.
1988, JGR, 93, 14383.
%
\bibitem[Aschwanden(2002)]{asc02} 
Aschwanden, M.~J.
2002, SSRv, 101, 1.
%
\bibitem[Batchelor(1960)]{Batchelor60} 
Batchelor, G.~K. 
1960, Cambridge University Press.
%
\bibitem[Bhattacharjee et al.(1998)]{bns98} 
Bhattacharjee, A., Ng, C. S., \& Spangler, S. R. 1998, ApJ, 494, 409.
%
\bibitem[Bieber et al.(1996)]{BieberEA96} 
Bieber, J. W., Wanner, W., \& Matthaeus, W. H. 
1996, JGR, 101, 2511.
%
\bibitem[Brown et al.(2009)]{bro09}
Brown, J.~C., Turkmani, R., Kontar, E.~P., MacKinnon, A.~L., \& Vlahos, L. 
2009, A\&A, 508, 993.
%
\bibitem[Browning \& Vekstein(2001)]{bv01}
Browning, P. K., \& Vekstein, G. E. 
2001, JGR, 106, 18677.
%
\bibitem[Bykov et al.(2008)]{byk08}
Bykov, A. M., Uvarov, Y. A., \& Ellison, D. C. 2008, \apjl, 689, L133
%
\bibitem[Cargill et al.(2012)]{cvb12}
Cargill, P. J., Vlahos, L., Baumann, G., Drake, J.~F., and Nordlund \AA
2012 SSRv., 173, 223.
%
\bibitem[Cowley(1978)]{Cowley78}                 
Cowley, S.~W.~H. 
1978, Planet. Space Sci., 26, 539. 
%
\bibitem[Dalena et al.(2012a)]{dal12a}
Dalena, S., Greco, A., Rappazzo, A.~F., Mace, R.~L., \& Matthaeus, W.~H.
2012, PhRvE, 86, 016402.
%
\bibitem[Dalena et al.(2012b)]{dal12b}
Dalena, S., Chuychai, P., Mace, R.~L., Greco, A., Qin, G., \& Matthaeus, W.~H.
2012, CoPhC, 183, 1974.
%
\bibitem[Dalla \& Browning(2005)]{DallaEA05} 
Dalla, S., \& Browning, P.~K. 
2005, A\&A, 436, 1103. 
%
\bibitem[de Jager(1986)]{jag86} 
de Jager, C. 
1986, SSRv, 44, 43.
%
\bibitem[Dmitruk et al.(1997)]{dg97} 
Dmitruk, P., \& G\'omez, D. O. 1997, ApJ, 484, L83
%
\bibitem[Dmitruk et al.(2003)]{dgm03} 
Dmitruk, P., G\'omez, D. O., \& Matthaeus, W. H. 2003, Phys. Plasmas, 10, 3584
%
\bibitem[Dmitruk et al.(2004)]{dmt04} 
Dmitruk, P., Matthaeus, W.~H., \& Seenu, N.
2004, ApJ, 617, 667.
%
\bibitem[Drake et al.(2005)]{DrakeEA05} 
Drake, J.~F., Shay, M.~A., Thongthai, W., \& Swisdak, M.
2005, PhRvL, 94, 095001. 
%
\bibitem[Drake et al.(2006)]{DrakeEA06} 
Drake, J.~F., Swisdak, M., Che, H., \& Shay, M.~A.
2006, Natur, 443, 553. 
%
\bibitem[Einaudi et al.(1996)]{evpp96}
Einaudi, G., Velli, M., Politano, H., \& Pouquet, A. 1996, ApJ, 457, L113
%
\bibitem[Emslie et al.(2003)]{ems03}
Emslie, A. G., Kontar, E. P., Krucker S., \& Lin, R. P. 
2003, ApJ, 595, L107.
%
\bibitem[Fletcher et al.(2007)]{fle07}
Fletcher, L., Hannah, I.~G., Hudson, H.~S., \& Metcalf, T.~R. 
2007, ApJ, 656, 1187.
%
\bibitem[Fraschetti \& Melia(2008)]{fras08}
Fraschetti, F., \& Melia, F. 2008, MNRAS, 391, 1100.
%
\bibitem[Fraschetti \& Giacalone(2012)]{fras12}
Fraschetti, F., \& Giacalone, J. 2012, \apj, 755, 114.
%
\bibitem[Frisch(1995)]{Frisch}
Frisch, U. 1995, Turbulence (Cambridge: Cambridge Univ. Press)
%
\bibitem[Giacalone \& Jokipii(1996)]{Giacalone&Jokipii96} 
Giacalone, J., \& Jokipii, J.~R.  
1996, JGR, 101, 11095.
%
\bibitem[Giacalone \& Jokipii(1999)]{Giacalone&Jokipii99} 
Giacalone, J., \& Jokipii, J.~R.  
1999, ApJ, 520, 204.
%
\bibitem[Giacalone(2005)]{Giacalone05}
Giacalone, J. 
2005, ApJ, 624, 765.
%
\bibitem[Giuliani et al.(2005)]{giu05}
Giuliani, P., Neukirch, T., \& Wood, P. 
2005, ApJ, 635, 636.
%
\bibitem[Gontikakis et al.(2013)]{GontiEA13}
Gontikakis, C., Patsourakos, S., Efthymiopoulos, C.,
Anastasiadis, A., \& Georgoulis, M.~K.
2013, \apj, 771, 126.
%
\bibitem[Gordovskyy et al.(2005)]{gor05} 
Gordovskyy, M., Zharkova, V. V., Voitenko, Yu. M., \& Goosens, M. 
2005, AdSpR, 35, 1743.
%
\bibitem[Gray et al.(1996)]{GrayEA96} 
Gray, P. C., Pontius, D. H., Jr., \& Matthaeus, W. H. 
1996, GRL, 23, 965.
%
\bibitem[Jokipii(1966)]{Jokipii66} 
Jokipii, J.~R.
1966, ApJ, 146, 480.
%
\bibitem[Kadomtsev \& Pogutse(1974)]{kp74} 
Kadomtsev, B. B., \& Pogutse, O. P. 
1974, JETP, 38, 283. 
%
\bibitem[Klein \& MacKinnon(2007)]{km07} 
Klein, K.~L. \& MacKinnon, A.~L.
2007, LNP, 725, 1.
%
\bibitem[Kobak \& Ostrowski(2000)]{KobakEA00}
Kobak, T., \& Ostrowski, M. 
2000, MNRAS {\bf 317}, 973.
%
\bibitem[Lee(1983)]{lee83} 
Lee, W.~W. 
1983, PhFl, 26, 556.
%
\bibitem[Lehe et al.(2009)]{LeheEA09} 
R. Lehe, I.~J. Parrish, and E. Quataert, 
2009, ApJ, 707, 409.
%
\bibitem[Litvinenko \& Somov(1993)]{LitvinenkoEA93} 
Litvinenko, Yu.~E. \& Somov, B.~V.
1993, SoPh, 146, 127.
%
\bibitem[Litvinenko(1996)]{Litvinenko96}
Litvinenko, Y.~E. 
1996, ApJ, 462, 997.
%
\bibitem[Mace et al.(2000)]{MaceEA00}
Mace, R.~L., Matthaeus, W.~H., \& Bieber, J.~W.
2000, ApJ, 538, 192.
%
\bibitem[Matthaeus et al.(1984)]{MatthaeusEA84} 
Matthaeus, W.~H., Ambrosiano, J.~J., \& Goldstein, M.~L.     
1984, PhRvL, 53, 1449.
%
\bibitem[Matthaeus \& Lamkin(1986)]{MatthaeusLamkin86}     
Matthaeus, W.~H., \& Lamkin, S.~L.
1986, PhFl, 29, 2513. 
%
\bibitem[Montgomery(1982)]{mon82} 
Montgomery, D. 
1982, PhST, 2, 83. 
%
\bibitem[Onofri et al.(2006)]{OnofriEA06} 
Onofri, M., Isliker, H., \& Vlahos, L. 
2006, PhRvL, 96, 151102. 
%
\bibitem[Oughton et al.(1994)]{OughtonEA94}
Oughton, S., Priest E.~R., \& Matthaeus, W.~H. 
1994, JFM., 280, 95.
%
\bibitem[Press et al.(1992)]{pre92} 
Press, W.~H., Teukolsky, S.~A., Vetterling, W.~T., Flannery, B. P.
1992, Cambridge University Press, 1992
%
\bibitem[Qin et al.(2002)]{qin02}
Qin, G., Matthaeus, W. H., and Bieber, J. W. 2002, \grl, 29, 7.
%
\bibitem[Rappazzo \& Parker(2013)]{rp13} 
Rappazzo, A. F., \& Parker, E. N.
2013, ApJL, 773, L2 
%
\bibitem[Rappazzo \& Velli(2011)]{rv11} 
Rappazzo, A. F., \& Velli, M.
2011, PhRvE, 83, 65401(R).
%
\bibitem[Rappazzo et al.(2007)]{rved07} 
Rappazzo, A. F., Velli, M., Einaudi, G., \& Dahlburg, R. B. 
2007, ApJL, 657, L47.
%
\bibitem[Rappazzo et al.(2008)]{rved08} 
Rappazzo, A. F., Velli, M., Einaudi, G., \& Dahlburg, R. B. 
2008, ApJ, 677, 1348. 
%
\bibitem[Rappazzo et al.(2012)]{rmr12}
Rappazzo, A.~F., Matthaeus, W.~H., Ruffolo, D., Servidio, S., \& Velli, M. 
2012, ApJL, 758, L14.
%
\bibitem[Ruffolo et al.(2003)]{RuffoloEA03} 
Ruffolo, D., Matthaeus, W.~H., \& Chuychai, P.
2003, ApJ, 597, L169. 
%
\bibitem[Smith at al.(2001)]{SmithEA01} 
Smith, C. W., Mullan, D. J., \& Ness, N. F. 
2001, JGR, 106, 18625.
%
\bibitem[Speiser(1965)]{Speiser65} 
Speiser, T.~W.  
1965, JGR, 70, 4219. 
%
\bibitem[Sonnerup(1971)]{Sonnerup71} 
Sonnerup, B.~U.~O. 
1971 JGR, 76, 8211. 
%
\bibitem[Strauss(1976)]{str76} 
Strauss, H. R. 
1976, PhFl, 19, 134.
%
\bibitem[Swann(1933)]{Swann33} 
Swann, W.~F.~G. 
1933, PhRv., 43, 217.
%
\bibitem[Takakura et al.(1995)]{tak95} 
Takakura, T., Kosugi, T., Sakao, T., et al. 
1995, PASJ, 47, 355.
%
\bibitem[Tu \& Marsch(1993)]{TuEA93} 
Tu, C. Y., \& Marsch, E.  
1993, JGR, 98, 1257.
%
\bibitem[Turkmani et al.(2005)]{TurkmaniEA05} 
Turkmani, R., Vlahos, L., Galsgaard, K., Cargill, P.~J., \& Isliker, H.
2005, ApJ, 620, L59.
%
\bibitem[Zank \& Matthaeus(1992)]{zm92}
Zank, G. P., \& Matthaeus, W. H. 1992, J. Plasma Phys., 48, 85.
\end{thebibliography}
\end{document}